\documentclass{osa-article}

\journal{oe}

\articletype{Research Article}

\usepackage{changes}
\definechangesauthor[color=red]{AB}
\definechangesauthor[color=blue]{CH}
\definechangesauthor[color=orange]{MP}

\usepackage{caption}
\usepackage{subcaption}

\begin{document}

\title{High power continuous laser at 461 nm based on a compact and high-efficiency frequency-doubling linear cavity}

\author{C.-H. Feng,\authormark{1, \dag} S. Vidal,\authormark{2, \dag} P. Robert,\authormark{1} P. Bouyer,\authormark{1} B. Desruelle,\authormark{3} M. Prevedelli,\authormark{4} J. Boullet,\authormark{2} G. Santarelli,\authormark{1} A. Bertoldi\authormark{1,*}}

\address{\authormark{1}LP2N, Laboratoire Photonique, Num\'erique et Nanosciences,
Universit\'e Bordeaux-IOGS-CNRS:UMR 5298, F-33400 Talence, France.\\
\authormark{2}ALPhANOV, Centre Technologique Optique et Laser, Rue Fran{\c c}ois Mitterrand, F-33400 Talence, France.\\
\authormark{3}$\mu$QuanS, Institut d'Optique d'Aquitaine, Rue Fran{\c c}ois Mitterrand, F-33400 Talence, France.\\
\authormark{4}Dipartimento di Fisica e Astronomia, Universit{\`a} di Bologna,
Via Berti-Pichat 6/2, I-40126 Bologna, Italy.\\
\authormark{$\dagger$}these authors contributed equally.\\
}
\email{\authormark{*}{\added[]{andrea.bertoldi@instituteoptique.fr}}}

\begin{abstract}
A Watt-level continuous and single frequency blue laser at 461 nm is obtained by frequency-doubling an amplified diode laser operating at 922 nm via a LBO crystal in a resonant Fabry-P\'{e}rot cavity. We achieved a best optical conversion efficiency equal to 87\% with more than 1 W output power in the blue, and limited by the available input power. The frequency-converted beam is characterized in terms of long term power stability, residual intensity noise, and geometrical shape. The blue beam has a linewidth of the order of 1 MHz, and we used it to magneto-optically trap $^{88}$Sr atoms on the 5s$^{2}\,^{1}$S$_0$ -- 5s5p$\,^{1}$P$_1$ transition. The low-finesse, linear-cavity doubling system is very robust, maintains the lock for several days, and is compatible with a tenfold increase of the power levels which could be obtained with fully-fibered amplifiers and large mode area fibers.
\end{abstract}

\section{Introduction}
Blue and violet laser light is increasingly adopted for both scientific and technical purposes, like in underwater communication \cite{Wu2017}, medical science \cite{Yun2017}, and quantum information\cite{Northup2014}. In the specific context of timekeeping, driven in the last years by the unrelenting improvements of optical lattice clocks, laser light in this band is typically required for the laser cooling of alkaline-earth and alkaline-earth-like atoms, e.g. for Yb at 399 nm, Ca at 422 nm, and Sr at 461 nm.

Different approaches are adopted to obtain the few hundreds of mW typically required for such cooling: laser diodes, a technologically simple solution, which can combine moderate power and narrow linewidth but requires optical injection of one or more slave modules \cite{Pagett2016}; single-pass second-harmonic (SH) generation with a bulk crystal \cite{Akamatsu2011}, which can achieve high power, while doubling the source linewidth but only with large input power; SH generation exploiting a travelling-wave cavity \cite{Woll1999,Leconte2018}, commonly folded in a bow-tie configuration for compactness. The latter configuration, which is widely adopted commercially \cite{toptica,leos}, offers high power and narrow linewidth at the price of a complex optical system and alignment procedure, a high sensitivity to vibration noise and, typically, astigmatism of the output beam.

\section{Frequency-doubling system}
In this article, we demonstrate high power CW operation at 461 nm. Our system exploits frequency doubling of an amplified infrared laser  in a linear cavity \cite{Klappauf2004} via a lithium triborate (LBO) crystal. Our solution combines low linear losses, ease of alignment, and high mechanical stability thanks to the reduced number of optical elements. Notably, we obtain an output power of 1 W, competitive with the state-of-the-art, and yet limited only by the available level of input power; the figure could be widely increased as soon as higher power will be available for the fundamental wavelength \cite{Su2021}.

\begin{figure}[htbp]
     \centering
     \begin{subfigure}[]{0.8\textwidth}
         \centering
         \includegraphics[width=0.92\textwidth]{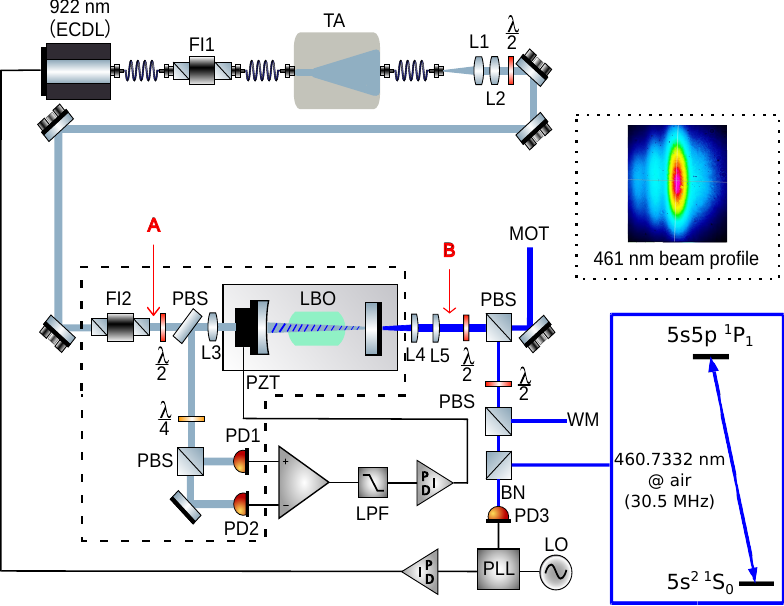}
         \caption{}
         \label{fig:schematic}
     \end{subfigure}
     \hfill
     \begin{subfigure}[]{0.8\textwidth}
         \centering
         \includegraphics[width=0.8\textwidth]{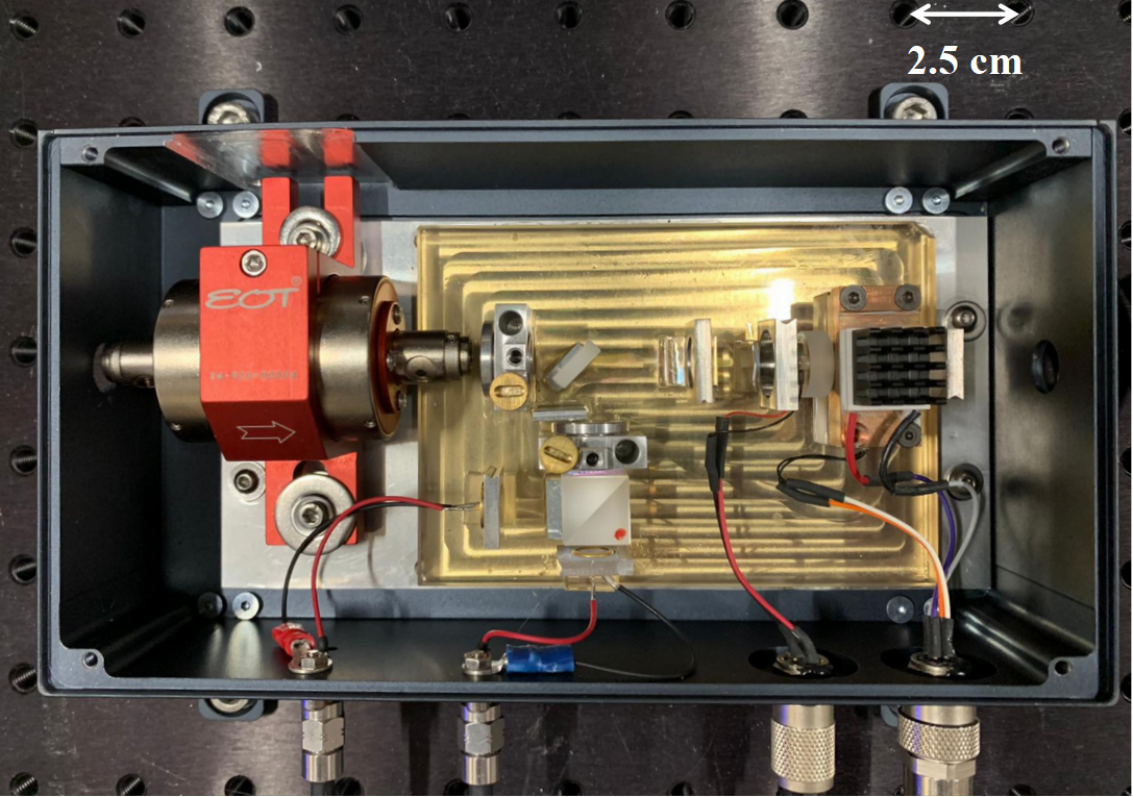}
         \caption{}
         \label{fig:picture}
     \end{subfigure}
     \caption{(a) Frequency doubling setup. The laser power at 922 nm injected to the cavity is measured at position A, the output power at 461 nm at position B. TA: tapered amplifier; FI1, FI2, FI3: Faraday isolators; BS: beam splitter; PBS: polarizing beam splitter; PD: photodetector; LPF: low-pass filter; L1, L2, L3: aspherical lenses; L4, L5: cylindrical lenses; PM: prism mirror; WM: wavelength meter. Inset: intensity profile of the output beam at 461 nm. (b) Top view of the elements within the dashed polygon in (a) and enclosed in an aluminum box to improve the thermal stability of the frequency doubling cavity.}
     \label{fig:SHG}
\end{figure}

The scheme of our frequency doubler is shown in Fig. \ref{fig:SHG}. The seed laser light at $\lambda$=922 nm and with a linewidth <200 kHz is generated by an ECDL (DL pro, Toptica) fiber coupled to a Tapered Amplifier (TA pro, Toptica). The system delivers a maximum output power of 2.2 W in free space, reduced to 1.36 W after coupling it to a single-mode fiber (P1-780PM-FC-1, Thorlabs) to clean the geometrical mode and ease its alignment on the doubling cavity (position A in Fig. \ref{fig:SHG}). 

Two $\approx$60 dB double-stage Faraday isolators (FI1 and FI2) decouple the TA from the the seed laser and from the doubling cavity respectively, thus protecting the system and avoiding frequency and intensity noise due to optical feedback.

We tested different kinds of non-linear crystals for the SH generation, as resumed in Tab. \ref{tab:material_prop}. Periodically-poled crystals operating at quasi-phase matching condition have been dismissed, despite a promising high conversion efficiency ($d_{\mathrm{eff}}$), because of their major limitations at high power induced by the photo-refractive beam distortion, the blue-induced infrared absorption \cite{Mabuchi1994}, and thermal effects \cite{Wei2017}. 
$\upbeta$-barium borate (BBO) \cite{Beier1997} and bismuth triborate (BIBO) \cite{Wen2016} crystals were discarded because of their large walk off angle (see Tab. \ref{tab:material_prop}). We opted for a LBO crystal from Crystal Laser characterised by a small walk off angle, high thermal conductivity, and an overall good long-term stability; its much smaller second-order non-linearity compared to periodically-poled crystals is compensated by a much higher damage threshold and crystal quality, a greater transparency at the operating wavelengths, and lower thermal lensing \cite{bass1995handbook, yang19916}.

\begin{table}[ht]
\centering
\resizebox{0.8\textwidth}{!}{
\begin{tabular}{c|c|c|c|c|c}
\hline
Parameters       & \shortstack{ppsLT \cite{Wei2017}} & \shortstack{ppKTP \cite{Kumar2009}} & BiBO \cite{Wen2016} & BBO \cite{Beier1997} & \shortstack{LBO \cite{Woll1999,Leconte2018,yang1991}}\\ \hline
\shortstack{Transparency \\ range (nm)} & 300-5000 & 350-4500 & 286-2500 &185-2600 & 160-2600 \\ \hline
Phase matching   & \shortstack{quasi-phase \\ matching \\ type I}  & \shortstack{quasi-phase \\ matching \\ type I} & critical  & critical & critical   \\ \hline
$d_{\rm{eff}}$ (pm/V) & 5 & 10.8 & 3.42 & 2 & 0.75 \\ \hline
\shortstack{Walk-off angle \\ (mrad)} & 0 & 0 & 44.5 & 61.34 & 9.38 \\ \hline
Thermal effects & yes & yes & no & no & no \\ \hline
\shortstack{Angular acceptance \\ (mrad$\times$cm)}  & na & \shortstack{55 ($\Delta \theta$) \\ 10 ($\Delta \phi$)} & 1.05 & 1.2 & 6.5 \\ \hline
\end{tabular}}
\caption{Key parameters for different kinds of crystals suitable to generate blue light by SH generation.}
\label{tab:material_prop}
\end{table}

The LBO crystal is tightly fixed on the cavity axis; it has a length of 15 mm and a cross section of 3$\times$3 mm$^2$. The crystal is cut for normal incidence $\theta=90 ^o$, $\phi=21^o$ at the phase matching at 48 $^oC$ and has a two layer antireflection coating working at 922 nm (reflectivity R<0.1\%) and 461 nm (R<0.3\%). The temperature of the LBO crystal is stabilized to the optimal phase-matching temperature of 48$^{\rm{o}}$C with a long term stability at the mK level, by controlling the temperature of the copper holder where it is fixed with thermally conductive glue. The temperature acceptance curve has a sinc$^2$ profile with a FWHM of $\sim3^{\rm{o}}$C for the given length of the crystal \cite{Lin1990}. The input mirror of the cavity is concave, has a nominal reflectivity at 922 nm equal to 98.8\% and a very high reflection coating at 461 nm; the output coupler is a flat mirror coated for high reflection at 922 nm and antireflection at 461 nm. The cavity is 18 mm long, and we used the Boyd-Kleinman equations \cite{Boyd1968} to optimize the conversion efficiency: this results in a radius of curvature of 15 mm for the input mirror, and in a fundamental cavity mode with a waist of $\sim$100 $\mu$m on the input mirror and $\sim$40 $\mu$m on the output one; the single-pass nonlinear conversion efficiency is evaluated to $\sim 1.0 \times 10^{-4}$ W$^{-1}$ \cite{Yariv1989}.

The amplified seed laser beam before the doubling cavity is mode-matched to the fundamental transverse mode of the cavity with a pair of lenses (L1 and L2), whose positions are precisely adjusted by means of micrometric translational stages. The polarization axis of the input beam is controlled with a $\lambda$/2 waveplate. To further protect the TA from optical feedback -- particularly strong when the cavity unlocks -- we placed an additional 30 dB optical isolator (FI3) resulting in an extra 12\% insertion loss at the input of the resonator.

The optical cavity is locked to the input seed laser using the modulation-free Hänsch-Couillaud stabilization technique \cite{hansch1980laser}; the cavity length is controlled by a piezo-electric device which translates one of the two mirrors defining the resonator with a 1 kHz bandwidth. The doubling efficiency is critically dependent on the mechanical and thermal stability of the system, hence great care was devoted to the design of the doubling cavity. Notably, mechanical robustness has been obtained by gluing the components of the doubling system to a Zerodur$^{\tiny{\textrm{\textregistered}}}$ plate and enclosing it in an aluminum box to passively improve its temperature stability.

\begin{figure}[htbp]
\centering
\includegraphics[width=0.8\linewidth]{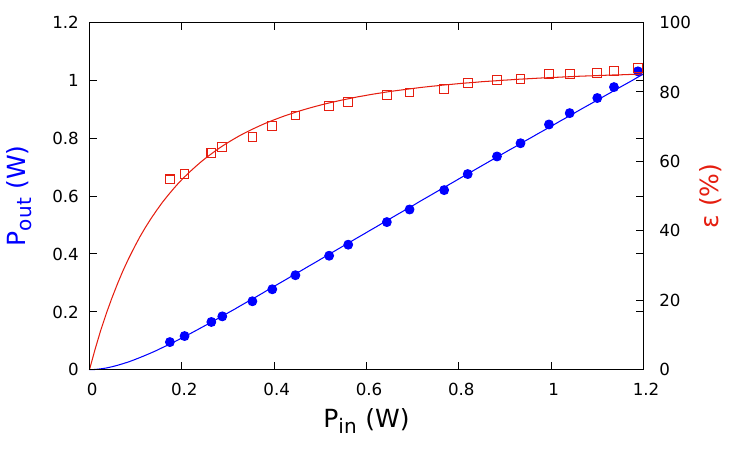}
\caption{Measured output power at 461 nm from the SH module (blue filled circles) and optical conversion efficiency versus incident fundamental laser power (red open squares). The solid lines result by fitting the experimental points with the theoretical curves.}
\label{fig:efficiency}
\end{figure}

\section{Characterization}
We measured the doubling cavity conversion efficiency $\epsilon \equiv P_{\rm{out}}/P_{\rm{in}}$, the power ratio between the SH output component at 461 nm and the fundamental input one at 922 nm, respectively measured at position B and A in Fig. \ref{fig:SHG}. The power of the SH component at 461 nm increases almost linearly with the fundamental laser power (blue circles in Fig. \ref{fig:efficiency}), indicating that the system works in the strong depletion regime dominated by non linear losses. The frequency conversion efficiency also increases, reaching a maximum value equal to 87\% with 1.05 W of blue light at 461 nm generated with 1.19 W at 922 nm (red squares in Fig. \ref{fig:efficiency}). In this configuration the intra-cavity power calculated using the single-pass conversion efficiency is $\sim$100 W; this value is in good agreement with what is obtained considering the build-up factor for an impedance-matched cavity. By fitting the experimental points with the expected theoretical behavior \cite{Yariv1989,Polzik1991} we: i.) confirm that thermal effects are negligible in our setup; ii.) obtain an input mirror reflectivity at 922 nm of 98.84(1)\%, which is very close to the nominal one, and estimate the linear losses in the cavity equal to 0.170(3)\%. From these values we infer a cavity finesse of $\sim 490$. Remarkably, the system performances in terms of both output power and conversion efficiency are presently limited by the available input laser power, which could be up-scaled for example exploiting fully-fibered solutions \cite{Rota2017}.

\begin{figure}[htbp]
\centering
\includegraphics[width=0.8\linewidth]{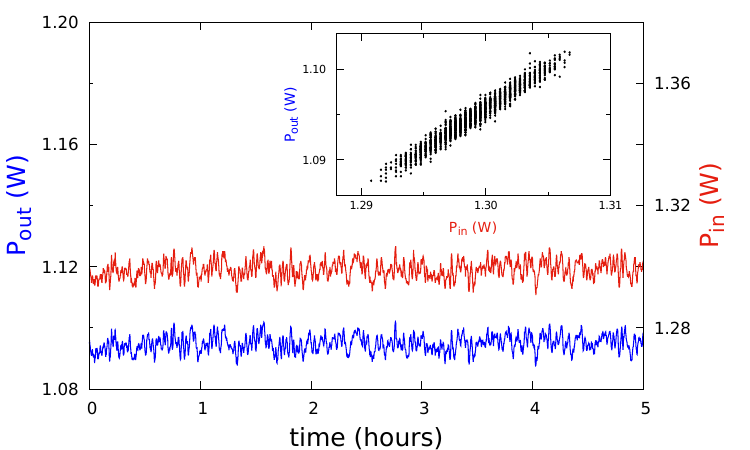}
\caption{Long-term intensity measurement of the SH output at 461 nm (blue curve, left y-axis) and incident fundamental laser at 922 nm (red curve, right y-axis). Inset: output intensity trace versus input intensity trace, demonstrating their strong correlation.}
\label{fig:output_461_922}
\end{figure}

The laser system can operate for a few days without any maintenance. The power stability of the frequency conversion system has been characterized by monitoring the SH power over several hours: the peak-to-peak fluctuations over 4 h are $\sim$1.2\% (see Fig. \ref{fig:output_461_922}). However, most fluctuations are strongly correlated with the varying input power (see the inset of Fig. \ref{fig:output_461_922}), and the uncorrelated relative fluctuations obtained after removing the linear dependence between P$_{\rm{in}}$ and P$_{\rm{out}}$ over the same time interval show an Allan deviation of $5\times 10^{-4}$ at 10 s decreasing to $7\times10^{-5}$ at 2500 s, with a standard deviation equal to $\sim 5.5 \times 10^{-4}$. This residual noise could be explained in terms of residual instabilities of the crystal temperature and of the cavity injection pointing. 

\begin{figure}[htbp]
\centering
\includegraphics[width=0.8\linewidth]{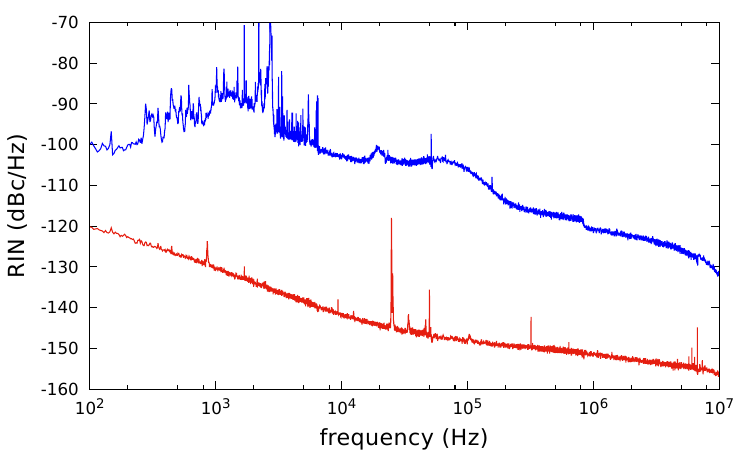}
\caption{Relative intensity noise of the fundamental light component (red curve) and of the SH component transmitted by the doubling cavity (blue curve).}
\label{fig:RIN}
\end{figure}

We measured the relative intensity noise (RIN) of both the fundamental and frequency converted component (see Fig. \ref{fig:RIN}). The spectrum of the fundamental light component starts at -120 dBc/Hz at 100 Hz, rolls-off as $f^{-1}$ till 20 kHz where the slope decreases to $f^{-1/3}$. The SH component shows an excess of noise in the acoustic band on a plateau at -105 dBc/Hz extending till 10 kHz, then it rolls off with a general $f^{-1}$ trend. The doubling process adds excess intensity noise in the whole Fourier frequency range, as already observed in previously published works based on resonant doubling cavities \cite{Eismann2016,Kwon2020} but not in single pass systems \cite{Dixneuf2021}.

The strong focusing of the fundamental mode on the crystal, imposed to obtain a high conversion efficiency, produces, as a side-effect, the multi-lobe cross-section of the frequency doubled beam, as shown in the inset of Fig. \ref{fig:SHG}-a. On the left side the intensity profile of the beam is well described by a sinc$^2$ function, which results from the windowing imposed by the crystal's angular acceptance (see Table 1, last row). Such structure is only partially found on the right side of the beam, because of the fundamental cavity mode being offset from the crystal axis, thus producing an asymmetric clipping of the frequency doubling process. To prevent similar problems, the cavity should be glued while monitoring its effective alignment \cite{Heinz2021}. The central lobe of the output beam contains more than 90\% of the power, and we correct its aspect ratio using two cylindrical lenses to improve the single mode fiber injection efficiency, which is >50\% of the total power.

To demonstrate the system potential in cold atom physics, we use it to implement a 3 dimensional magneto-optical trap (MOT) for $^{88}$Sr atoms, capturing the pre-cooled atomic flux produced by a compact 2D MOT similar to that reported in \cite{Lamporesi2013,Nosske2017}. The frequency of the SH component is referenced to the Sr spectroscopy via a frequency-shift optical lock with a red detuning $\Delta=1.5 \; \Gamma$ with respect to the blue cooling transition 5s$^{2}\,^{1}$S$_0$ -- 5s5p$\,^{1}$P$_1$ transition of $^{88}$Sr ($\Gamma \sim 2 \pi \times 30.5$ MHz is the transition linewidth) \cite{Appel2009}. The resulting linewidth of the servo-controlled blue laser is of the order of one MHz, hence significantly below that of the cooling transition. We use the beam to produce the 6 cooling beams for the 3D MOT with integrated fiber splitters, and obtain in this way a stable operation of the trap with $\sim 10^9$ atoms.

\section{Conclusion}
In conclusion, we realized a robust frequency doubled system based on a LBO crystal placed in a linear Fabry-P\'{e}rot cavity; the system delivers more than 1 W of power at 461 nm with an efficiency of 87\%, limited by the available power on the fundamental mode. A higher output power and more integrated setup could exploit a fully-fibered amplifier \cite{Rota2017} and large mode area fibers \cite{Corre2021} for the fundamental frequency. \medskip

\textit{Note:} Another scheme to generate Watt-level blue light by frequency doubling in a cavity a vertical-external-cavity surface-emitting laser (VECSEL) has been reported recently \cite{tinsley2021watt}. \medskip

\noindent
{\bf Acknowledgments.}
We thank Hanyu Ye for splicing an optoisolator to the seed laser and Clément Dixneuf for helping with the data acquisition for the RIN measurement.
\\
{\bf Funding.}
This work was partly supported by the Conseil R{\'e}gional d'Aquitaine
(grant IASIG-3D and grant USOFF), the ``Agence Nationale pour la
Recherche" (grant EOSBECMR \# ANR-18-CE91-0003-01, grant ALCALINF \#
ANR-16-CE30-0002-01), the European Union (EU) (FET-Open project
CRYST$^3$), Horizon 2020 QuantERA ERA-NET (grant TAIOL \#
ANR-18-QUAN-00L5-02) and LAPHIA-IdEx Bordeaux. P. R. thanks DGA for
financial support.
\\
{\bf Disclosures.} SV and JB: ALPhANOV, F-33400 Talence, France (E), BD: $\mu$Quans, F-33400 Talence, France (E,I), PB: $\mu$Quans, F-33400 Talence, France (I). The authors declare no other conflicts of interest.
\\
{\bf Data availability.}
Data relative to the results presented in this paper may be obtained from the authors upon reasonable request.

%\bibliography{main}

\end{document}